\documentclass[conference]{IEEEtran}
\usepackage{amssymb}
\usepackage{amsmath}
\usepackage{cite}
\usepackage{url}
\usepackage{xcolor}
\usepackage{cite,graphicx,amsmath,amssymb}
\usepackage{subfigure}
\usepackage{citesort}
\usepackage{fancyhdr}
\usepackage{mdwmath}
\usepackage{mdwtab}
\usepackage{caption}
\usepackage{amsthm}
\usepackage{algorithm}
\usepackage{algorithmic}
\usepackage{makecell}
\usepackage{diagbox}
\allowdisplaybreaks
\usepackage[
top    = 1.70cm,
bottom = 1.05in,
left   = 0.63 in,
right  = 0.63 in]{geometry}

\newtheorem{remark}{Remark}
\newtheorem{theorem}{Theorem}

\newtheorem{lemma}{Lemma}

\newtheorem{corollary}{Corollary}

\begin{document}
\title{Semantic Communication-assisted Physical Layer Security over Fading Wiretap Channels}


\author{\IEEEauthorblockN{Xidong Mu\IEEEauthorrefmark{1} and Yuanwei Liu\IEEEauthorrefmark{1}}
\IEEEauthorblockA{\IEEEauthorrefmark{1}School of Electronic Engineering and Computer Science, Queen Mary University of London, U.K.\\E-mail:\{xidong.mu, yuanwei.liu\}@qmul.ac.uk}}

\maketitle
\begin{abstract}
A novel semantic communication (SC)-assisted secrecy transmission framework is proposed. In particular, the legitimate transmitter (Tx) sends the superimposed semantic and bit stream to the legitimate receiver (Rx), where the information may be eavesdropped by the malicious node (EVE). As the EVE merely has the conventional bit-oriented communication structure, the semantic signal acts as the type of beneficial \emph{information-bearing artificial noise (AN)}, which not only keeps strictly confidential to the EVE but also interferes with the EVE. The ergodic (equivalent) secrecy rate over fading wiretap channels is maximized by jointly optimizing the transmit power, semantic-bit power splitting ratio, and the successive interference cancellation decoding order at the Tx, subject to both the instantaneous peak and long-term average power constraints. To address this non-convex problem, both the optimal and suboptimal algorithms are developed by employing the Lagrangian dual method and the successive convex approximation method, respectively. Numerical results show that the proposed SC-assisted secrecy transmission scheme can significantly enhance the physical layer security compared to the baselines using the conventional bit-oriented communication and no-information-bearing AN. It also shows that the proposed suboptimal algorithm can achieve a near-optimal performance.

\end{abstract}
\section{Introduction}
Recently, semantic communication (SC) has been proposed as a new paradigm for information transmission in 6G~\cite{9955525,Qin}. Compared to conventional information transmission under the Shannon paradigm (termed as bit-oriented communication), where the total source is converted into bit sequences for transmission, SC only focuses on the key part of the source which is relevant to the specific meaning/actions/goals~\cite{Qin}. By doing so, on the one hand, the original source can be significantly compressed before transmission and the required wireless resources can be greatly reduced. On the other hand, compared to bit-oriented communication, SC makes the receiver 'understand' the information for realizing specific tasks. This is promising to be employed in the ``human-to-machine'' and ``machine-to-machine'' communication scenarios.\\
\indent Motivated by the above advantages, extensive research efforts have been devoted to developing SC approaches with the aid of machine learning tools. On the one hand, efficient SC approaches have been conceived for conventional text/speech/image/video transmission. For example, the authors of \cite{9398576} employed the deep learning-based semantic text transmission tool, namely DeepSC. It showed that the performance of DeepSC can significantly outperform conventional bit communication within the low and moderate signal-to-noise ratio (SNR) ranges. Based on the DeepSC-enabled text transmission, the authors of \cite{9763856} proposed the semantic rate performance metric for quantifying the performance of SC. Moreover, the authors of \cite{8723589} developed a joint source-channel coding approach for image transmission. The authors of \cite{du2023ai} further proposed a full-duplex device-to-device SC approach to reduce the computation tasks in the mixed reality application. On the other hand, researchers also developed SC approaches for new emerging tasks. For example, the authors of \cite{9796572} proposed the SC-based image transmission for enabling unmanned aerial vehicles to classify images. Furthermore, the authors of \cite{9998551} studied the timeliness of information in task-oriented communications.\\
\indent It can be observed that SC provides a new transmission strategy for future wireless networks. To this end, some initial works began to investigate the interplay between SC and other physical layer techniques to further improve communication performance. For example, the authors of \cite{10158994} proposed an opportunistic semantic and bit-oriented communication scheme for controlling the co-channel interference in uplink non-orthogonal multiple access (NOMA) systems. The authors of \cite{du2023semantic} developed an inverse semantic-aware wireless sensing framework with the aid of reconfigurable intelligent surface (RIS) technologies.\\
\indent Against the above background, we explore the potential of SC in physical layer security (PLS). In particular, we propose a novel SC-assisted secrecy transmission framework over fading wiretap channels, where one legitimate transmitter (Tx) sends information to one legitimate receiver (Rx) under the eavesdropping of one malicious node (EVE). Both semantic and bit-oriented encoder/decoder are equipped at the legitimate Tx and Rx, while only the conventional bit-oriented encoder/decoder is equipped at the EVE. The main idea of the proposed SC-assisted secrecy transmission framework is that the Tx sends the superimposed semantic and bit streams to the Rx to deliver the intended information, where the semantic stream not only keeps strictly confidential to the EVE but also can interfere with the EVE when eavesdropping the bit stream, i.e., acting as one type of beneficial artificial noise (AN). Based on the proposed framework, we maximize the ergodic (equivalent) secrecy rate by jointly optimizing the transmit power, power splitting ratio among the two streams, and the successive interference cancellation (SIC) decoding order of the Tx, subject to the peak and average power constraints. To solve this non-convex optimization problem, both optimal and suboptimal algorithms are proposed by employing the Lagrangian dual method and the successive convex approximation (SCA) method, respectively. Our numerical results show that the proposed SC-assisted secrecy transmission scheme can significantly improve the secrecy rate compared to the baselines using conventional bit-oriented communication and no-information-bearing AN. Moreover, the proposed suboptimal algorithm can achieve near-optimal performance. 

\section{System Model and Problem Formulation}
As shown in Fig. \ref{model}, we investigate a fundamental three-node single-input-single-output (SISO) secrecy transmission over fading wiretap channels, which consists of one legitimate Tx, one legitimate Rx, and one malicious EVE. All nodes are assumed to have a single antenna. In particular, the legitimate Tx and Rx are assumed to be equipped with both the new semantic-channel encoder/decoder and the conventional bit-oriented source-channel encoder/decoder, while the malicious EVE is assumed to only have the conventional bit encoder/decoder. The quasi-static block fading channel model is assumed for both the legitimate Tx-Rx link and the wiretapping Tx-EVE link. The instantaneous Tx-Rx and Tx-EVE channel coefficients at the fading state $v$ are denoted by ${h_L}\left( v \right)$ and ${h_E}\left( v \right)$, respectively. The channel coefficients are assumed to be unchanged during each fading state $v$ and independently vary between different fading states. To characterize the maximum performance gain and explore useful insights, in this work, we assume that all channel coefficients are perfectly known at the Tx. 
\subsection{SC-assisted Secrecy Transmission}
\begin{figure}[!t]
  \centering
  \includegraphics[width=3in]{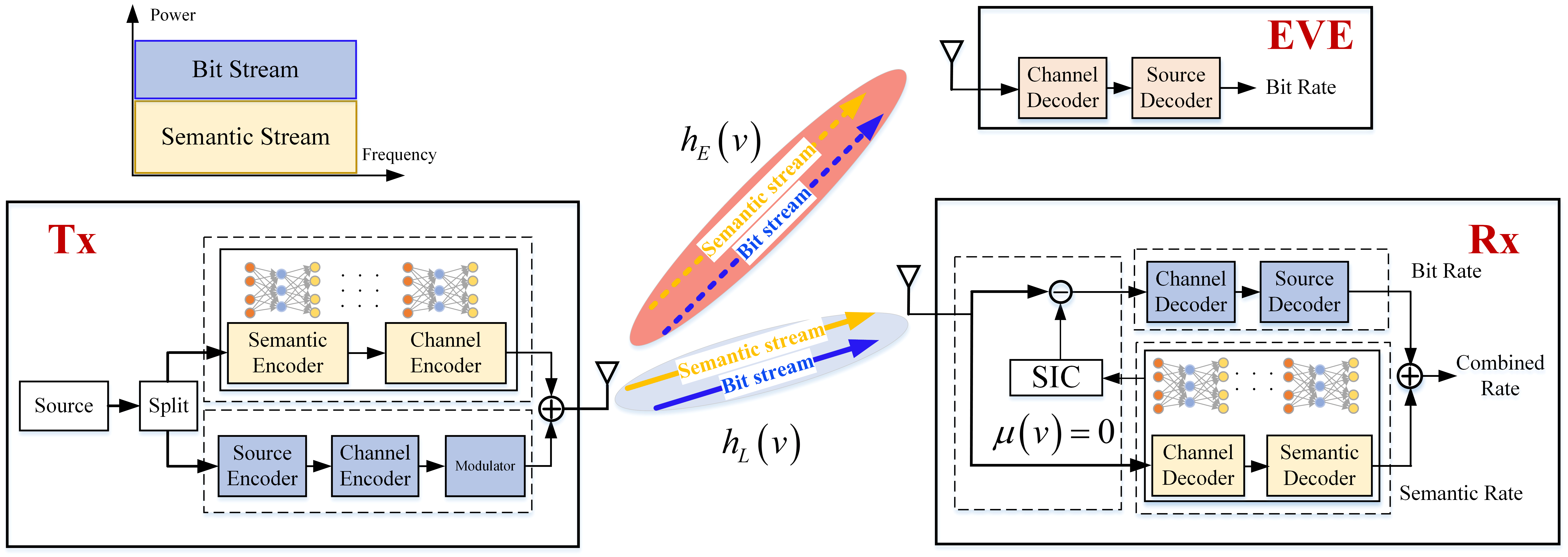}\\
  \caption{The proposed SC-assisted secrecy transmission framework.}\label{model}
\end{figure}

As illustrated, we propose a novel SC-assisted secrecy transmission framework. Recalling the fact that the legitimate Tx/Rx has both semantic and bit-oriented communication structures, as illustrated in Fig. \ref{model}, the original source at the Tx is firstly split into two parts, where one part is processed by the semantic-channel encoder (termed as a semantic stream) and the other part is processed by the conventional bit-oriented source-channel encoder (termed as a bit stream). Let $x_s$ and $x_b$ denote the normalized semantic symbol from the semantic-channel encoder and the normalized information symbol from the conventional bit-oriented source-channel encoder, respectively. Therefore, the superimposed semantic and bit signal transmitted by the Tx at fading state $v$ can be expressed as
\begin{equation}\label{transmitted}
x\left( v \right) = \sqrt {\beta \left( v \right)p\left( v \right)} {x_s} + \sqrt {\left( {1 - \beta \left( v \right)} \right)p\left( v \right)} {x_b},
\end{equation}
where ${p\left( v \right)}$ denotes the employed instantaneous transmit power of the Tx at fading state $v$, and ${0 \le \beta \left( v \right) \le 1}$ denotes the corresponding power portion allocated to the semantic stream. Here, we consider both the peak power constraint (PPC) and the average power constraint (APC) at the Tx. For PPC, we have $\left\{ {p\left( v \right) \le \widehat P,\forall v} \right\}$, where ${\hat P}$ represents the maximum instantaneous transmit power that can be used at the Tx for each fading state $v$. For APC, we have ${{\mathbb{E}}_v}\left[{p\left( v \right)} \right] \le \overline P$, where ${{\mathbb{E}}_v}\left[  \cdot  \right]$ is the expectation operation over $v$ and $\overline P$ represents the maximum long-term average transmit power that can be used at the Tx over the entire fading states. Without loss of generality, we have $\overline P \le \widehat P$. 

Accordingly, the received signal at the legitimate Rx and the malicious EVE are respectively given by
\begin{subequations}\label{received}
\begin{align}
\label{L}&\begin{array}{l}
  {y_L}\left( v \right)\\
   = {h_L}\left( v \right)\left( {\sqrt {\beta \left( v \right)p\left( v \right)} {x_s} + \sqrt {\left( {1 - \beta \left( v \right)} \right)p\left( v \right)} {x_b}} \right) + {z_L},
  \end{array}\\
\label{E}&\begin{array}{l}
  {y_E}\left( v \right)\\
   = {h_E}\left( v \right)\left( {\sqrt {\beta \left( v \right)p\left( v \right)} {x_s} + \sqrt {\left( {1 - \beta \left( v \right)} \right)p\left( v \right)} {x_b}} \right) + {z_E},
  \end{array}
\end{align}
\end{subequations}
where ${z_L} \sim {\cal C}{\cal N}\left( {0,\sigma _L^2} \right)$ and ${z_E} \sim {\cal C}{\cal N}\left( {0,\sigma _E^2} \right)$ denote the additive white Gaussian noise (AWGN) at the legitimate Rx and the malicious EVE, respectively.

For the legitimate Rx, both the received semantic and bit signals, $\left\{ {{x_s},{x_b}} \right\}$, are desired and can be decoded. However, for the malicious EVE only having the conventional bit-oriented source-channel decoder, only the bit signal, $\left\{ {{x_b}} \right\}$ can be decoded for wiretapping. In the following, we introduce the communication performance of the legitimate Rx and the malicious EVE.

\subsubsection{Achievable Communication Rate at the Rx} The legitimate Rx successively decodes the received superimposed signal with the aid of SIC, i.e., the Rx first decodes one type of signal and then subtracts it from the received signal, before decoding the other type of signal. Then, the SNR or signal-to-interference-plus-noise (SINR) when decoding the bit signal, $x_b$, and semantic signal, $x_s$, are respectively given by
\begin{subequations}\label{SINR}
  \begin{align}
  \label{B}&{\gamma _{L,b}\left( v \right)} = \frac{{\left( {1 - \beta \left( v \right)} \right)p\left( v \right){g_L}\left( v \right)}}{{\mu \left( v \right)\beta \left( v \right)p\left( v \right){g_L}\left( v \right) + 1}},\\
  \label{S}&{\gamma _{L,s}\left( v \right)} = \frac{{\beta \left( v \right)p\left( v \right){g_L}\left( v \right)}}{{\left( {1 - \mu \left( v \right)} \right)\left( {1 - \beta \left( v \right)} \right)p\left( v \right){g_L}\left( v \right) + 1}},
  \end{align}
\end{subequations}
where ${g_L}\left( v \right) \buildrel \Delta \over = \frac{{{{\left| {{h_L}\left( v \right)} \right|}^2}}}{{\sigma _L^2}}$. The binary variable $\mu \left( v \right) \in \left\{ {0,1} \right\}$ denotes the SIC decoding order employed at the Rx for fading state $v$. If the bit siganl is decoded first, we have $\mu \left( v \right) = 1$; Otherwise, $\mu \left( v \right) = 0$. An example of $\mu \left( v \right) = 0$, i.e., the decoding order from semantic signal to bit signal, is illustrated in Fig. \ref{model}.

Accordingly, the achievable bit rate (bit/s/Hz) for decoding $x_b$ is given by 
\begin{align}\label{Brate}
{R_b}\left( v \right) = {\log _2}\left( {1 + {\gamma _{L,b}}\left( v \right)} \right).
\end{align}

To quantify the communication performance of SC, we employ the performance metric, namely semantic rate, proposed in \cite{9763856} for DeepSC-enabled semantic text transmission \cite{9398576}. Assuming that, on average, each sentence at the Tx contains $L$ words, representing $I$ semantic information (measured in semantic units (suts)). Additionally, each word generates an average of $K \in {Z}^{+}$ semantic symbols through DeepSC. According to \cite{9763856}, the achievable semantic rate (suts/s/Hz) for decoding $x_s$ is given by:  
\begin{align}\label{Srate}
{S_s}\left( v \right) = \frac{I}{{KL}}\varepsilon_K \left( {{\gamma _{L,s}}\left( v \right)} \right),
\end{align}
where $0 \le \varepsilon_K \left( {{\gamma _{L,s}}\left( v \right)} \right) \le 1$ denotes the semantic similarity function which quantifies the semantic loss between the original text source and the recovered text source in SC. Its value depends on the employed DeepSC scheme, $K$, and the received SNR/SINR, ${\gamma _{L,s}}\left( v \right)$. In \cite{9763856}, $ \varepsilon$ can only be experimentally obtained by using the DeepSC tool. To provide a closed-form expression of $ \varepsilon$ for facilitating theoretical study, in our previous work \cite{9953095}, we approximate $ \varepsilon$ for any given $K$ with high accuracy using a generalized logistic function as follows:
\begin{align}\label{logistic}
\varepsilon_K \left( {\gamma _{L,s}}\left( v \right) \right) \approx  {A_{K,1}} + \frac{{{A_{K,2}} - {A_{K,1}}}}{{1 + {e^{ - \left( {{C_{K,1}}10\lg \left( {{\gamma _{L,s}}\left( v \right)} \right) + {C_{K,2}}} \right)}}}}.
\end{align}
Here, for different $K$, the lower (left) asymptote and the upper (right) asymptote of the generalized logistic function are denoted by ${A_{K,1}}>0$ and ${A_{K,2}}>0$, respectively. The corresponding logistic growth rate and mid-point are represented by ${{C_{K,1}}}>0$ and ${{C_{K,2}}}$, respectively. 

As illustrated in Fig. \ref{model}, after decoding the two types of signal, Rx will combine the decoded results. Then, the achievable communication rate at Rx for fading state $v$ should be ${R_b}\left( v \right)+{{S_s}\left( v \right)}$. However, the units of the two rates are different, so we transform the semantic rate into the equivalent bit rate (bit/s/Hz) as follows \cite{9763856}:
\begin{align}\label{Brate2}
{R_s}\left( v \right) = \frac{{\rho L}}{I}{S_s}\left( v \right) = \frac{\rho }{K}\varepsilon_K \left( {{\gamma _{L,s}}\left( v \right)} \right),
\end{align}
where $\rho=40$ (bits/word) represents the average number of bits required for transmitting one word if ASCII code is used for bit-oriented source-channel encoder/decoder and the semantic similarity achieved by bit transmission is assumed to be 1 \cite{9763856}. Based on the above equivalent transformation, the achievable communication rate at the Rx for fading state $v$ after combing the two streams can be expressed as
\begin{align}\label{Total}
{R_L}\left( v \right) = {R_b}\left( v \right) + {R_s}\left( v \right).
\end{align}

\subsubsection{Achievable Communication Rate at the EVE} As the EVE can only decode the bit signal, $x_b$, by treating the semantic signal, $x_s$, as interference, the corresponding received SINR at fading state $v$ can be expressed as
\begin{align}\label{E_SINR}
{\gamma _E}\left( v \right) = \frac{{\left( {1 - \beta \left( v \right)} \right)p\left( v \right){g_E}\left( v \right)}}{{\beta \left( v \right)p\left( v \right){g_E}\left( v \right) + 1}},
\end{align}
where ${g_E}\left( v \right) \buildrel \Delta \over = \frac{{{{\left| {{h_E}\left( v \right)} \right|}^2}}}{{\sigma _E^2}}$. Then, the achievable communication rate of EVE at fading state $v$ is ${R_E}\left( v \right) = {\log _2}\left( {1 + {\gamma _E}\left( v \right)} \right)$.

\subsubsection{Secrecy Communication Rate} Based on the above discussion, the achievable secrecy communication rate at fading state $v$ is given by
\begin{align}\label{secrecy}
{\cal R}\left( {p \left( v \right),\beta \left( v \right),\mu\left( v \right)} \right) = {\left[ {{R_L}\left( v \right) - {R_E}\left( v \right)} \right]^ + },
\end{align}
where ${\left[  \cdot  \right]^ + } \buildrel \Delta \over = \max \left( { \cdot ,0} \right)$. Therefore, the ergodic secrecy rate is ${\cal E}={{\mathbb{E}}_v}\left[  {\cal R}\left( {p \left( v \right),\beta \left( v \right),\mu\left( v \right)} \right)  \right]$.

\begin{remark}\label{difference}
\emph{Since the EVE only has the capability of decoding the conventional bit stream, the semantic stream in the proposed scheme can act as a beneficial information-bearing AN. On the one hand, the semantic signal conveys useful information to the Rx, see \eqref{S}. On the other hand, the semantic signal is strictly confidential to the EVE and can interfere with the EVE during the decoding of bit signal, see \eqref{E_SINR}. }
\end{remark}

\subsection{Problem Formulation}
In this paper, we aim to maximize the ergodic secrecy rate, ${\cal E}$, by jointly optimizing the transmit power, $\left\{ {p\left( v \right),\forall v} \right\}$, the power allocation factor, $\left\{ {\beta\left( v \right),\forall v} \right\}$, and the SIC decoding order, $\left\{ {\mu \left( v \right),\forall v} \right\}$, of each fading state, subject to the PPC and APC. We can formulate the studied optimization problem as follows:
\begin{subequations}\label{Problem}
\begin{align}
\mathop {\max }\limits_{\left\{ {p \left( v \right),\beta \left( v \right),\mu\left( v \right)} \right\}} &\;{{\mathbb{E}}_v}\left[ {{\cal R}\left( {p \left( v \right),\beta \left( v \right),\mu\left( v \right)} \right)} \right]\\
\label{C11}{\rm{s.t.}}\;\;&{{\mathbb{E}}_v}\left[ {p\left( v \right)} \right] \le \overline P,\\
\label{C12}&0 \le p\left( v \right) \le \widehat P, \forall v,\\
\label{C13}&0 \le \alpha \left( v \right) \le 1, \forall v,\\
\label{C14}&{\mu\left( v \right) \in \left\{ {0,1} \right\},\forall v}.
\end{align}
\end{subequations}
As the objective function is non-concave and the integer SIC decoding order constraint \eqref{C14} is non-convex, problem \eqref{Problem} is a non-convex optimization problem. In the following, we will propose both optimal and suboptimal solutions to problem \eqref{Problem}.

\section{Proposed Solutions}
\subsection{Optimal Solution to Problem \eqref{Problem}}
To optimally solve problem \eqref{Problem}, we first show that it satisfies the ``time-sharing'' condition \cite{timeshare} with the following lemma.
\begin{lemma}\label{TS}
\emph{For problem \eqref{Problem} given the APCs $\overline P_a$ and $\overline P_b$, the corresponding optimal solutions are denoted by $\left\{ {{p_a}\left( v \right),{\beta _a}\left( v \right),{\mu _a}\left( v \right)} \right\}$ and $\left\{ {{p_b}\left( v \right),{\beta _b}\left( v \right),{\mu _b}\left( v \right)} \right\}$, respectively. Then, for any $0 \le \theta  \le 1$, there always exists a feasible solution $\left\{ {{p_c}\left( v \right),{\beta _c}\left( v \right),{\mu _c}\left( v \right)} \right\}$ such that}
\[{\cal E}_c \ge \theta {\cal E}_a  + \left( {1 - \theta } \right){\cal E}_b, \]
\[{{\mathbb{E}}_v}\left[{{p_c}\left( v \right)} \right] \le \theta {\overline P _a} + \left( {1 - \theta } \right){\overline P _b},\]
\emph{where ${\cal E}_i={{\mathbb{E}}_v}\left[  {\cal R}\left( {p_i \left( v \right),\beta_i \left( v \right),\mu_i\left( v \right)} \right)  \right]$ is calculated by substituting the given solutions into \eqref{secrecy}, $i \in \left\{ {a,b,c} \right\}$.}
\begin{proof}
{Considering that each fading state $v$ has a certain amount of time, we can construct a feasible solution $\left\{ {{p_c}\left( v \right),{\beta _c}\left( v \right),{\mu _c}\left( v \right)} \right\}$ by allocating a $\theta$ percentage of the time and the remaining $1-\theta$ percentage of the time to solutions $\left\{ {{p_a}\left( v \right),{\beta _a}\left( v \right),{\mu _a}\left( v \right)} \right\}$ and $\left\{ {{p_b}\left( v \right),{\beta _b}\left( v \right),{\mu _b}\left( v \right)} \right\}$, respectively. By combining all fading states with the above constructed solution, we have ${{\mathbb{E}}_v}\left[ {{{\cal R}}\left( {p_c \left( v \right),\beta_c \left( v \right),\mu_c\left( v \right)} \right)} \right] ={\cal E}_c=\theta {\cal E}_a  + \left( {1 - \theta } \right){\cal E}_b= \theta {{\mathbb{E}}_v}\left[ {{{\cal R}}\left( {p_a \left( v \right),\beta_a \left( v \right),\mu_a\left( v \right)} \right)} \right]  + \left( {1 - \theta } \right){{\mathbb{E}}_v}\left[ {{{\cal R}}\left( {p_b \left( v \right),\beta_b \left( v \right),\mu_b\left( v \right)} \right)} \right]$ and ${{\mathbb{E}}_v}\left[{{p_c}\left( v \right)} \right] = \theta {{\mathbb{E}}_v}\left[{{p_a}\left( v \right)} \right] + \left( {1 - \theta } \right){{\mathbb{E}}_v}\left[{{p_c}\left( v \right)} \right] \le \theta {\overline P _a} + \left( {1 - \theta } \right){\overline P _c}$. Therefore, the proof of Lemma 1 is completed.}
\end{proof}
\end{lemma}
\textbf{Lemma 1} implies that the ``time-sharing'' condition \cite{timeshare} holds for problem \eqref{Problem}. It means that there is zero duality gap between the primal problem \eqref{Problem} and the Lagrange dual problem, i.e., strong duality holds~\cite{convex}. As a result, we can employ the Lagrange duality method to optimally solve problem \eqref{Problem}. The Lagrangian of problem \eqref{Problem} can be expressed as
\begin{equation}\label{Lagrangian2}
  {{{\mathcal{L}}}}\left( {p \left( v \right),\beta \left( v \right),\mu\left( v \right),\lambda } \right)  = {{\mathbb{E}}_v}\left[ {{{\cal R}_c}\left( v \right)}   \right] + \lambda \left\{ {\overline P  - {{\mathbb{E}}_v}\left[ {p\left( v \right)} \right]} \right\}, 
\end{equation}
where the non-negative Lagrange multiplier $\lambda$ is associated with the APC \eqref{C11}. As a result, the partial Lagrange dual function of problem \eqref{Problem} is given by
\begin{equation}\label{Lagrangian_dual2}
\begin{gathered}
{g}\left( {\lambda} \right) = \hfill \\
\mathop {\max }\limits_{0 \le p\left( v \right) \le \widehat P,0 \le \beta \left( v \right) \le 1,\mu \left( v \right) \in \left\{ {0,1} \right\},\forall v} {{{\mathcal{L}}}}\left( {p \left( v \right),\beta \left( v \right),\mu \left( v \right),\lambda } \right).\hfill \\
\end{gathered}
\end{equation}
Note that we can decompose problem \eqref{Lagrangian_dual2} into several parallel subproblems sharing the same structure. Each subproblem corresponds to one distinct fading state. In the following discussion, we drop the fading state index $v$ for brevity. Under any given $\lambda$, the associated subproblem for a particular fading state can be rewritten as follows:
\begin{equation}\label{sub2}
\mathop {\max }\limits_{0 \le p \le \hat P, 0 \le \beta  \le 1,\mu  \in \left\{ {0,1} \right\}} {\overline {\mathcal{L}} }\left( {p, \beta, \mu } \right),
\end{equation}
where ${\overline {{\mathcal{L}}} }\left( {p ,\alpha , \mu} \right) =  {\cal R}\left( {p, \beta, \mu } \right) - \lambda p$. To solve problem \eqref{sub2}, we need to compare the respective maximum value associated with the cases of $\mu=1$ and $\mu=0$, which yields the two subproblems \eqref{U1} and \eqref{U2} as shown at the top of the next page. Before solving the two problems, we first have the following lemma.
\begin{figure*}[!t]
\normalsize
\begin{equation}\label{U1}
{\mathcal{F}}_{1}\left( {p_1^*,\beta _1^*} \right) = \mathop {\max }\limits_{0 \le p \le \hat P,0 \le \beta  \le 1} {\left[ {\frac{\rho }{K}{\varepsilon _K}\left( {\beta p{g_L}} \right) + {{\log }_2}\left( {1 + \frac{{\left( {1 - \beta } \right)p{g_L}}}{{\beta p{g_L} + 1}}} \right) - {{\log }_2}\left( {1 + \frac{{\left( {1 - \beta } \right)p{g_E}}}{{\beta p{g_E} + 1}}} \right)} \right]^ + } - \lambda p,
\end{equation}
\begin{equation}\label{U2}
{\mathcal{F}}_0 \left( {p_0^*,\beta _0^*} \right) = \mathop {\max }\limits_{0 \le p \le \hat P,0 \le \beta  \le 1} {\left[ {\frac{\rho }{K}{\varepsilon _K}\left( {\frac{{\beta p{g_L}}}{{\left( {1 - \beta } \right)p{g_L} + 1}}} \right) + {{\log }_2}\left( {1 + {\left( {1 - \beta } \right)p{g_L}} } \right) - {{\log }_2}\left( {1 + \frac{{\left( {1 - \beta } \right)p{g_E}}}{{\beta p{g_E} + 1}}} \right)} \right]^ + } - \lambda p,
\end{equation}
\end{figure*}
\begin{lemma}\label{order}
\emph{When $\frac{{1}}{{{g_L}}} \ge \frac{{1}}{{{g_E}}}$, the optimal decoding order to problem \eqref{sub2} is ${\mu ^*} = 0$.}
\begin{proof}
{When $\frac{{\sigma _L^2}}{{{g_L}}} \ge \frac{{\sigma _E^2}}{{{g_E}}}$, the value of ${{{\log }_2}\left( {1 + \frac{{\left( {1 - \beta } \right)p{g_L}}}{{\beta p{g_L} + 1}}} \right) - {{\log }_2}\left( {1 + \frac{{\left( {1 - \beta } \right)p{g_E}}}{{\beta p{g_E} + 1}}} \right)}$ is always negative under any given $0<p\le \hat P$ and $0\le \beta<1$. In this case, the optimal power splitting ratio to problem \eqref{U1} when ${\mu} = 1$ is $\beta _1^* = 1$, i.e., not allocating power to the bit stream. It can be verified that ${{\cal F}_1}\left( {p_1^*,\beta _1^* = 1} \right) = {{\cal F}_0}\left( {p_0^*,\beta _0^* = 1} \right) \le {{\cal F}_0}\left( {p_0^*,\beta _0^*} \right)$. Therefore, the proof of Lemma 2 is completed.}
\end{proof}
\end{lemma}
Therefore, the optimal solutions to problem \eqref{sub2} are given By
\begin{equation}\label{solutions}
\left\{ \begin{array}{l}
{\mu ^*} = 1,\;p = p_1^*,\beta  = \beta _1^*,\;\;{\rm{if}}\;\frac{{1}}{{{g_L}}} < \frac{{1}}{{{g_E}}}\;\;{\rm{and}}\;{{\cal F}_1} > {{\cal F}_0},\\
{\mu ^*} = 0,\;p = p_0^*,\beta  = \beta _0^*,\;{\rm{otherwise}},
\end{array} \right.
\end{equation}
where $\left( {p_1^*,\beta _1^*} \right)$ and $\left( {p_0^*,\beta _0^*} \right)$ are obtained from \eqref{U1} and \eqref{U2} via exhaustive search.

Based on the above discussion, for any given $\lambda$, problem \eqref{Lagrangian_dual2} can be solved by solving problem \eqref{sub2} for different fading states. Recalling the fact of the strong duality holds for problem \eqref{Problem}, problem \eqref{Problem} can be optimally solved by iteratively solving problem \eqref{Lagrangian_dual2} with fixed $\lambda$ and update $\lambda$ with the bisection method until the PPC \eqref{C11} is satisfied with equality.

\subsection{Suboptimal Solution to Problem \eqref{Problem}}
Note that the exhaustive search has to be employed for each fading state to find the optimal solution to problem \eqref{Lagrangian_dual2}, which leads to potential high computational complexity. To this end, we propose a successive convex approximation (SCA)-based suboptimal algorithm to solve problem \eqref{Lagrangian_dual2} with a lower computational complexity. 

\textbf{Lemma 2} implies that there is a higher probability that the optimal SIC decoding order is ${\mu  = 0}$. Inspired by this, to handle the integer constraint caused by the SIC decoding order in problem \eqref{Lagrangian_dual2}, we fix $\left\{ {\mu \left( v \right) = 0,\forall v} \right\}$ for all fading states. Let us define ${p_s}\left( v \right) \buildrel \Delta \over = \beta \left( v \right)p\left( v \right)$, ${p_b}\left( v \right) \buildrel \Delta \over = \left( {1 - \beta \left( v \right)} \right)p\left( v \right),\forall v$, and introduce the below auxiliary variables $\left\{ {\chi \left( v \right),\forall v} \right\}$ such that 
\begin{equation}\label{auxiliary}
\chi \left( v \right) = {A_{K,1}} + \frac{{{A_{K,2}} - {A_{K,1}}}}{{1 + {e^{ - \left( {{C_{K,1}}10\lg \left( {\frac{{{p_s}\left( v \right){g_L}\left( v \right)}}{{{p_b}\left( v \right){g_L}\left( v \right) + 1}}} \right) + {C_{K,2}}} \right)}}}}.
\end{equation}
With the above definition, problem \eqref{Lagrangian_dual2} can be re-expressed as follows:
\begin{subequations}\label{Problem2}
\begin{align}
&\mathop {\max }\limits_{\left\{ {{p_s}\left( v \right),{p_b}\left( v \right),\chi \left( v \right)} \right\}} {{\mathbb{E}}_v}\left[ {{R_L}\left( {\chi \left( v \right),{p_b}\left( v \right)} \right) - {R_E}\left( {{p_b}\left( v \right),{p_s}\left( v \right)} \right)} \right]\\
\label{C21}&{\rm{s.t.}}\;\;\chi \left( v \right) \le {A_{K,1}} + \frac{{{A_{K,2}} - {A_{K,1}}}}{{1 + {e^{ - \left( {{C_{K,1}}10\lg \left( {\frac{{{p_s}\left( v \right){g_L}\left( v \right)}}{{{p_b}\left( v \right){g_L}\left( v \right) + 1}}} \right) + {C_{K,2}}} \right)}}}},\\
\label{C22}&\;\;\;\;\;\;\;\;{{\mathbb{E}}_v}\left[ {p_s\left( v \right)+p_b\left( v \right)} \right] \le \overline P,\\
\label{C23}&\;\;\;\;\;\;\;\;0 \le p_i\left( v \right) \le \widehat P, \forall i \in \left\{ {s,b} \right\}, \forall v,
\end{align}
\end{subequations}
where ${R_L}\left( {\chi \left( v \right),{p_b}\left( v \right)} \right) \buildrel \Delta \over = \frac{\rho }{K}\chi \left( v \right) + {\log _2}\left( {1 + {p_b}\left( v \right){g_L}\left( v \right)} \right)$ and ${R_E}\left( {{p_b}\left( v \right),{p_s}\left( v \right)} \right) \buildrel \Delta \over = {\log _2}\left( {1 + \frac{{{p_b}\left( v \right){g_E}\left( v \right)}}{{{p_s}\left( v \right){g_E}\left( v \right) + 1}}} \right)$. Problem \eqref{Problem2} is a non-convex optimization problem due to the non-convex function ${R_E}\left( {{p_b}\left( v \right),{p_s}\left( v \right)} \right)$ and the non-convex constraint \eqref{C21}. Note that ${R_E}\left( {{p_b}\left( v \right),{p_s}\left( v \right)} \right)$ can be rewritten as the difference of two concave functions, and the convex upper bound by employing the first-order Taylor expansion at given local points $\left\{ {p_s^{\left( r \right)}\left( v \right),p_b^{\left( r \right)}\left( v \right)} \right\}$ in the $r$th iteration is given by \eqref{U3}, as shown at the top of the next page.
\begin{figure*}[!t]
\normalsize
\begin{equation}\label{U3}
\begin{array}{l}
{R_E}\left( {{p_b}\left( v \right),{p_s}\left( v \right)} \right) = {\log _2}\left( {1 + {p_s}\left( v \right){g_E}\left( v \right) + {p_b}\left( v \right){g_E}\left( v \right)} \right) - {\log _2}\left( {1 + {p_s}\left( v \right){g_E}\left( v \right)} \right) \le \overline R _E^{\left( r \right)}\left( {{p_b}\left( v \right),{p_s}\left( v \right)} \right)\\
\buildrel \Delta \over = {\log _2}\left( {1 + p_s^{\left( r \right)}\left( v \right){g_E}\left( v \right) + p_b^{\left( r \right)}\left( v \right){g_E}\left( v \right)} \right) + \frac{{{g_E}\left( v \right)\left( {{p_s}\left( v \right) - p_s^{\left( r \right)}\left( v \right) + {p_b}\left( v \right) - p_b^{\left( r \right)}\left( v \right)} \right)}}{{\left( {1 + p_s^{\left( r \right)}\left( v \right){g_E}\left( v \right) + p_b^{\left( r \right)}\left( v \right){g_E}\left( v \right)} \right)\ln 2}} - {\log _2}\left( {1 + {p_s}\left( v \right){g_E}\left( v \right)} \right) .
\end{array}
\end{equation}
\begin{equation}\label{U4}
\begin{array}{l}
\ln \left( {\chi \left( v \right) - {A_{K,1}}} \right) + {C_{K,1}}10\lg \left( {{p_b}\left( v \right){g_L}\left( v \right) + 1} \right) \le {C_{K,1}}10\lg \left( {{p_s}\left( v \right){g_L}\left( v \right)} \right) + {C_{K,2}} + \ln \left( {{A_{K,2}} - \chi \left( v \right)} \right).
\end{array}
\end{equation}
\hrulefill \vspace*{0pt}
\end{figure*}

For the non-convex constraint \eqref{C21}, we first rewrite it into \eqref{U4}, as shown at the top of the next page. The left-hand-side of \eqref{U4} is non-convex but concave with respect to ${\chi \left( v \right)}$ and ${{p_b}\left( v \right)}$. By employing the first-order Taylor expansion at given local points $\left\{ {{\chi ^{\left( r \right)}}\left( v \right),p_b^{\left( r \right)}\left( v \right)} \right\}$ in the $r$th iteration, an upper bound is given by
\begin{equation}\label{eta}
\begin{array}{l}
{\eta ^{\left( r \right)}}\left( {\chi \left( v \right),{p_b}\left( v \right)} \right) \buildrel \Delta \over =  \ln \left( {{\chi ^{\left( r \right)}}\left( v \right) - {A_{K,1}}} \right) + \frac{{\chi \left( v \right) - {\chi ^{\left( r \right)}}\left( v \right)}}{{{\chi ^{\left( r \right)}}\left( v \right) - {A_{K,1}}}}\\
 + {C_{K,1}}10\lg \left( {1 + p_b^{\left( r \right)}\left( v \right){g_L}\left( v \right)} \right) + \frac{{10{C_{K,1}}{g_L}\left( v \right)\left( {{p_b}\left( v \right) - p_b^{\left( r \right)}\left( v \right)} \right)}}{{\left( {1 + p_b^{\left( r \right)}\left( v \right){g_L}\left( v \right)} \right)\ln 10}}.
\end{array}
\end{equation}

Employing \eqref{U3}-\eqref{eta}, problem \eqref{Problem2} can be rewritten as follows:
\begin{subequations}\label{Problem3}
\begin{align}
&\mathop {\max }\limits_{\left\{ {{p_s}\left( v \right),{p_b}\left( v \right),\chi \left( v \right)} \right\}} {{\mathbb{E}}_v}\left[ {{R_L}\left( {\chi \left( v \right),{p_b}\left( v \right)} \right) - \overline R _E^{\left( r \right)}\left( {{p_b}\left( v \right),{p_s}\left( v \right)} \right)} \right]\\
\label{C31}&{\rm{s.t.}}\;\;\begin{array}{l}
  {\eta ^{\left( r \right)}}\left( {\chi \left( v \right),{p_b}\left( v \right)} \right)\\
   \le {C_{K,1}}10\lg \left( {{p_s}\left( v \right){g_L}\left( v \right)} \right) + {C_{K,2}} + \ln \left( {{A_{K,2}} - \chi \left( v \right)} \right),
  \end{array}\\
\label{C32}&\;\;\;\;\;\;\;\;\;\;\eqref{C22},\eqref{C23}.
\end{align}
\end{subequations}
It can be found that problem \eqref{Problem3} is convex, which can be efficiently solved with convex optimization software, such as CVX \cite{cvx}. Therefore, the original non-convex problem \eqref{Problem2} can be solved by iteratively solving the convex problem \eqref{Problem3} with the local points $\left\{ {p_s^{\left( r \right)}\left( v \right),p_b^{\left( r \right)}\left( v \right),{\chi ^{\left( r \right)}}\left( v \right)} \right\}$ until convergence, thus obtaining a locally optimal solution. In particular, the solutions obtained in $r$th iteration are used as the input local points for the $\left( {r + 1} \right)$th iteration. Note that only a convex optimization problem \eqref{Problem3} needs to be solved in each iteration, the proposed suboptimal algorithm has a polynomial computational complexity in the worst case.

\section{Numerical Results}
In this section, we present numerical results to validate the effectiveness of the proposed SC-assisted secrecy transmission scheme as well as the performance of the developed optimal and suboptimal algorithms. Let $d_L$ and $d_E$ denote the Tx-Rx distance and the Tx-EVE distance, respectively. The large-scale distance-dependent path loss is modelled as $PL  = {{PL}_0}{\left( {{1 \mathord{\left/
{\vphantom {1 d_i}} \right.
\kern-\nulldelimiterspace} d_i}} \right)^\alpha }$, where ${{PL}_0}=-30$ dB denotes the reference path loss at 1 meter, $\alpha =4$ denotes the path loss exponent, and $d_i, \forall i \in \left\{ {L,E} \right\}$ denotes the corresponding link distance in meters. For the small-scale fading, we assume that the Tx-Rx and Tx-EVE channels at each fading state follow
independent and identically distributed Rayleigh fading. In the simulations, ${d_L} = {d_E} = 30$ meter. The PPC is set to $\widehat P= 10$ W and the noise power is set to $\sigma_L^2 =\sigma_E^2 = - 80$ dBm. 

\begin{figure}[!t]
  \centering
  \includegraphics[width=2.8in]{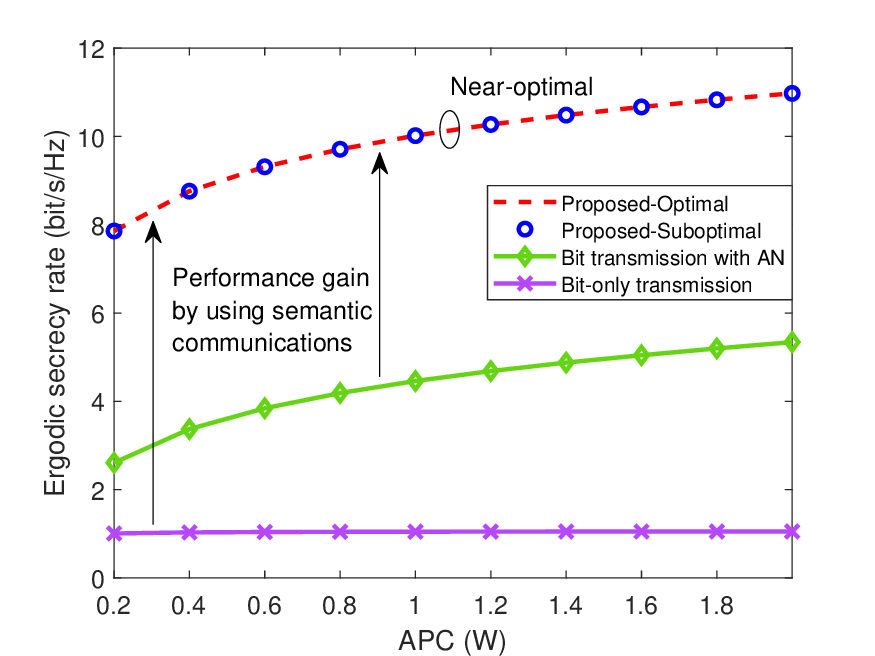}\\
  \caption{Ergodic secrecy rate of different transmission schemes versus the APC.}\label{figure1}
\end{figure}

\begin{figure}[!t]
  \centering
  \includegraphics[width=2.8in]{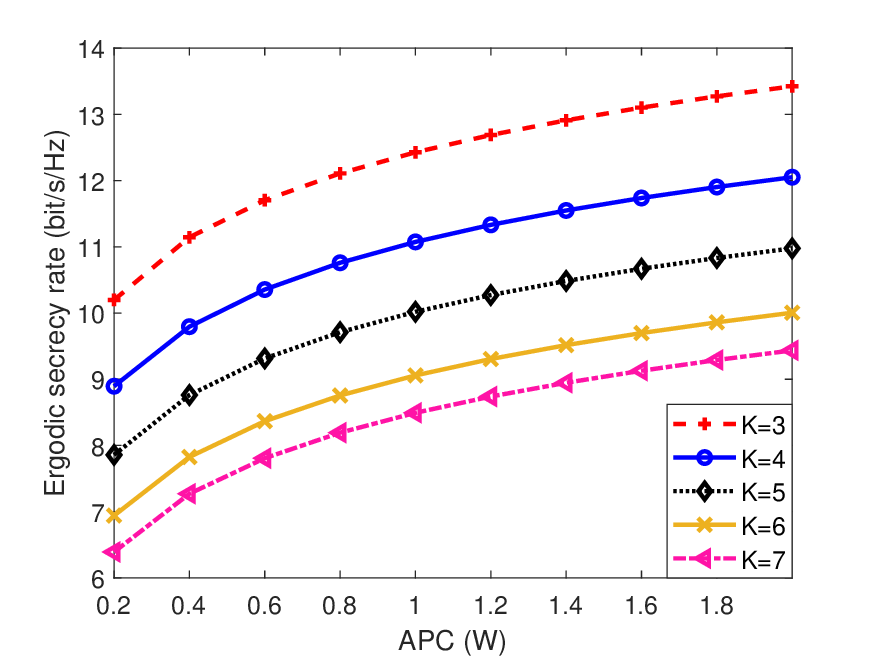}\\
  \caption{Ergodic secrecy rate of different semantic encoder schemes versus the APC.}\label{figure2}
\end{figure}

For performance comparison, we consider the following two benchmark schemes based on the conventional bit-oriented transmission. (1) \textbf{Bit-only transmission}: In this scheme, the Tx only employs the conventional bit-oriented transmission and all the information can be eavesdropped by the EVE. The corresponding secrecy communication rate at fading state $v$ can be expressed as ${{\cal R}_{{\rm{bit\_only}}}}\left( v \right) = {\left[ {{{\log }_2}\left( {1 + \frac{{p\left( v \right){g_L}\left( v \right)}}{{\sigma _L^2}}} \right) - {{\log }_2}\left( {1 + \frac{{p\left( v \right){g_E}\left( v \right)}}{{\sigma _E^2}}} \right)} \right]^ + }$. The resulting ergodic secrecy rate maximization problem can be solved by using the proposed algorithm by setting $\left\{ {\beta \left( v \right) = 0,\forall v} \right\}$. (2) \textbf{Bit transmission with no-information-bearing AN~\cite{5524086}}: In this scheme, the Tx sends both the bit and no-information-bearing AN signals. In particular, we assume that the AN signal can be cancelled by the Rx but cannot be cancelled by the EVE. The corresponding secrecy communication rate at fading state $v$ can be expressed as ${{\cal R}_{{\rm{bit\_AN}}}}\left( v \right) = {\left[ {{{\log }_2}\!\left( {1 \! + \! \frac{{\left( {1 - \beta \left( v \right)} \right)p\left( v \right){g_L}\left( v \right)}}{{\sigma _L^2}}} \right) \!\!-\! {{\log }_2}\!\left( {1 \! + \! \frac{{\left( {1 - \beta \left( v \right)} \right)p\left( v \right){g_E}\left( v \right)}}{{\beta \left( v \right)p\left( v \right){g_E}\left( v \right) + \sigma _E^2}}} \right)} \right]^ + }$. The resulting ergodic secrecy rate maximization problem can be solved by using the proposed algorithm by setting $\left\{ {\mu \left( v \right) = 0, {R_s}\left( v \right) = 0, \forall v} \right\}$.\\
\indent In Fig. \ref{figure1}, we compare the ergodic secrecy rate achieved by different transmission schemes. For the employed SC, we set $K=5$ and the corresponding parameters for the generalized logistic function are ${A_{K,1}}=0.37$, ${A_{K,2}}=0.98$, ${{C_{K,1}}}=0.2525$, and ${{C_{K,2}}}=-0.7895$. It can be observed that the ergodic secrecy rate increases with the APC, $\overline P$, except for the bit-only transmission scheme. This underscores the importance of employing jamming schemes to guarantee the PLS. Moreover, compared to the conventional bit transmission with AN scheme, the proposed SC-assisted secrecy transmission scheme can achieve a significant performance gain. This is because the semantic signal in the proposed scheme can be regarded as a type of beneficial AN signal, which not only interferes with the EVE but also delivers useful information to the Rx. The above performance comparison confirms the effectiveness of the proposed SC-assisted secrecy transmission scheme. Furthermore, it can be seen from Fig. \ref{figure1} that the performance gap between the optimal and suboptimal solutions is negligible, i.e., the proposed suboptimal algorithm can achieve near-optimal performance. \\
\indent In Fig. \ref{figure2}, we investigate the impact of the semantic encoding scheme, $K$, on the achieved ergodic secrecy rate of the proposed SC-assisted secrecy transmission scheme. All the results in Fig. \ref{figure2} are obtained by the proposed suboptimal algorithm. It can be observed that the achieved ergodic secrecy rate decreases with $K$. This is because equation \eqref{Brate2} implies that a higher $K$ employed in the SC leads to a lower equivalent bit rate, i.e., a lower performance gain of SC over conventional bit-oriented communication. The above results underscore the importance of developing an efficient semantic encoder/decoder, which can use less number of semantic symbols to deliver the desired information. 

\section{Conclusions}
An SC-assisted secrecy transmission framework was proposed for enhancing the physical layer security in future wireless networks. The legitimate Tx communicates with the legitimate Rx using the superimposed semantic and bit steams under the presence of the EVE. Due to the bit-information-only decoding capability of the EVE, the semantic stream was a beneficial AN to deliver the confidential information to the Rx and interfere with the EVE. The transmit power, power splitting ratio between the two streams, and the SIC decoding order at the Tx were joint optimized to maximize the ergodic (equivalent) secrecy rate over the wiretap fading channels, subject to the PPC and APC. To solve this problem, Lagrangian dual method-based and SCA-based algorithms were developed to obtain the optimal and suboptimal solutions. Our numerical results showed that the developed suboptimal achieves near-optimal performance and the proposed SC-assisted secrecy transmission can achieve significant secrecy communication gain.

\bibliographystyle{IEEEtran}
\bibliography{mybib}
\end{document}